\documentclass[sigconf]{acmart}

\usepackage{subfigure}
\usepackage{multicol}
\usepackage[ruled,vlined]{algorithm2e}
\usepackage{newtxmath}
\usepackage{array}
\usepackage{url}

\setcopyright{none}\acmConference[]{ACM}{2020}{} 
\copyrightyear{2020}
\acmYear{2020}
\acmBooktitle{ss}

\begin{document}
\fancyhead{}

\title{MarioMix: Creating Aligned Playstyles for Bots with Interactive Reinforcement Learning}


\author{Christian Arzate Cruz}
\affiliation{%
  \institution{The University of Tokyo}
  \city{Tokyo}
  \country{Japan}
  }
\email{arzate.christian@gmail.com}

\author{Takeo Igarashi}
\affiliation{%
  \institution{The University of Tokyo}
  \city{Tokyo}
  \country{Japan}
}
\email{takeo@acm.org}

\renewcommand{\shortauthors}{Arzate and Igarashi}

\begin{abstract}
  In this paper, we propose a generic framework that enables game developers without knowledge of machine learning to create bot behaviors with playstyles that align with their preferences. Our framework is based on interactive reinforcement learning (RL), and we used it to create a behavior authoring tool called MarioMix. This tool enables non-experts to create bots with varied playstyles for the game titled \textit{Super Mario Bros.} The main interaction procedure of MarioMix consists of presenting short clips of gameplay displaying precomputed bots with different playstyles to end-users. Then, end-users can select the bot with the playstyle that behaves as intended. We evaluated MarioMix by incorporating input from game designers working in the industry.
\end{abstract}



\maketitle

\section{Introduction}
\label{sec::Introduction}
Research on artificial intelligence (AI) for games is experiencing an era of rapid advances in many different areas. However, these novel approaches are rarely adopted by the game industry; academic research on AI for games (game AI) and AI development in commercial games are focused on different problems \cite{Yannakakis2015Panorama}. While academic researchers aim to create automatic AI techniques for general proposes, game designers prefer robust techniques that perform as intended in a particular application. 

For instance, creating bots with varied playstyles can be useful for game designers in early development stages, such as playtesting using bots \cite{Ariyurek2019, Pfau2018}. Unfortunately, such bot behaviors have not been widely adopted in the game industry. This is because creating bots with varied playstyles is time-consuming and often requires end-users to have knowledge of machine learning \cite{Hohman2018, Yang2019, Greydanus2018}. Furthermore, designing a playstyle that aligns with the game designer's intentions is a major challenge in the AI research field \cite{Leike2018}. In this paper, we propose a generic framework based on interactive reinforcement learning (RL) \cite{Fails2003, Amershi2014, Sacha2016} that will enable the design of authoring tools for bot behaviors with playstyles that align with the intention of end-users who have no knowledge of RL.

The interactive RL setting involves a human-in-the-loop that gives feedback to the underlying RL algorithm to improve its performance or guide it to solve a problem in a particular manner \cite{Amershi2014, Leike2018}. One major challenge in interactive RL is making it usable in high-dimension environments since most interactive RL-based applications require up to millions of human feedback samples to get good results \cite{Arzate2020, Griffith2013, Li2013, Rosenfeld2018}. To overcome the sample inefficiency of current interactive RL methods, our framework proposes a novel interaction procedure that consists of mixing pre-computed policies and a search method to find playstyles that fit the end-users' preferences.  

The mixture of policies is performed by assigning different policies (each policy exhibits a different playstyle) to specific segments of a game level. A similar interaction is proposed in \cite{Sorensen2016}, where users can guide an interactive evolutionary computation (IEC) \cite{Takagi2001}-based algorithm by deciding which individuals (or behavior) are used for breeding in the next iteration of the evolutionary algorithm. The main difference between the IEC-based algorithm in \cite{Sorensen2016} and our work is that we can achieve short interaction cycles because we follow a top-down method that only requires users to evaluate short gameplay clips and select their favorite. Furthermore, the computation of behaviors is done offline; end-users do not need to wait for computations to be performed while they use a tool based on our generic framework.

As a proof-of-concept of this framework, we build a tool designed around it called MarioMix. The aim of MarioMix is to create aligned playstyles for bots in a clone of the \textit{Super Mario Bros.} video game. Finally, we evaluated MarioMix via a user study in which professional game developers working in the industry participated. The results of our study reveal that MarioMix is an effective tool for creating aligned bot behaviors through interactive RL in a game industry context.

\section{Related Work}
\label{sec::RelatedWork}
Most research on ML focuses on algorithms that automatically find a solution for a particular problem. However, we find iML \cite{Fails2003} incorporates a human-in-the-loop that integrates domain knowledge to enhance the solution of the model at hand. In this paper, we focus on interactive RL-based approaches for agent alignment and refer our readers interested in agent alignment to the survey on this subject in \cite{Leike2018}. Thus, our research scope includes works that use interactive RL to generate agent behaviors that characterize the end-user's intentions.


\subsection{Interactive Reinforcement Learning}
There is extensive research on interactive reinforcement learning (RL) \cite{Arakawa2018, Rosenfeld2018, Elizalde2012, Korpan2017, Wang2018, McGregor2017, Griffith2013, Li2013, Amir2016, Fachantidis2017}. For a survey on interactive RL, please refer to \cite{Arzate2020}. 

Arguably, one of the most common approaches in interactive RL is policy shaping (PS) \cite{Griffith2013}. The basic idea of PS is to augment the policy of an RL model using human feedback. This feedback can be an actionable advice \cite{Macglashan2017, Krening2018} or a critique \cite{Griffith2013, Krening2018}. In actionable advice, users provide feedback to the model on what they think is the optimal action to perform in a particular state. A critique consists of binary feedback that indicates whether an action performed by the bot at a given state is good. 

In contrast, our framework uses a PS method but at a higher level since we mix policies, and each of them activates at a predefined space in the game level. This PS approach has the advantage of fast interaction cycles, which is essential to avoid causing users to become fatigued. Furthermore, by using precomputed policies, we enable the use of interactive RL in a complex environment, such as \textit{Super Mario Bros}.

There are alternative ways to provide actionable advice or feedback to an agent, such as using apprentice learning \cite{Abbeel2004}, inverse RL \cite{Ng1999, Ziebart2008, Ziebart2009}, or demonstration \cite{Taylor2011}. All these techniques utilize users' demonstrations of how a bot is supposed to perform a certain task as input. 
 
Our framework also uses human feedback in the form of gameplay demonstrations in a similar manner as \cite{Taylor2011, Ng1999, Ziebart2008, Ziebart2009, Abbeel2004}. However, unlike these approaches, our framework uses gameplay demonstrations to search within the playstyle space. Thus, we characterize the users' playstyle, and then we present them the bot with the most similar playstyle to theirs in our behavior dataset. This interaction channel is aimed at reducing the needed time and feedback for the alignment problem \cite{Griffith2013, Li2013, Rosenfeld2018}.

\subsection{AI-assisted Tools for Agent Behavior}
AI tools have been used extensively for procedural content generation (PCG) \cite{Summerville2018, Roberts2015, Khalifa2016, Sheffield2018}. However, the application of AI-assisted tools for agent behavior is still underexploited. Most approaches focus on the automatic generation of agent behavior; for a survey on this subject, refer to \cite{Yannakakis2015Panorama}. 

For instance, the NERO platform has the objective of generating believable behaviors for a first-person shooter \cite{Stanley2005}. This neuroevolution-based approach has the particularity of improving the performance of its agents by allowing imitation learning; the agent learns from demonstrating a policy that mimics an expert's trajectory \cite{Abbeel2004, Finn2016, Ho2016}. Another example can be found in \cite{Sorensen2016}, where an interactive evolutionary computation (IEC) approach \cite{Takagi2001} is implemented. In this application, users have the task of guiding an evolutionary algorithm by selecting which bots will either be used for breeding in the next iteration or die. 

In our generic framework, we use interactive RL instead of IEC, as in the previously mentioned works. Similar to the NERO platform \cite{Stanley2005}, we use gameplay trajectories as input, but we use them as a search method in the playstyle space. The interaction procedure in MarioMix is similar to the work in \cite{Sorensen2016} since we also present the user clips of different playstyle behaviors. However, we use the interaction procedure to generate a mixture of policies with varied playstyles that align with user preferences.

\section{Proposed Approach}
\label{sec::Approach}
The components of our generic framework are as follows: a problem expressed as a Markov decision process (MDP), either flat or hierarchical; an algorithm that solves the MDP(s); a dataset of policies, each with a distinctive playstyle; a search function for the playstyle space; a playstyle similarity metric; and a method for delimiting the activation of the policies.

\subsection{MDP Solver}
To create a policy dataset, we use a model-based reinforcement learning (RL) architecture called HRLB\textasciicircum2 \cite{Arzate2018}. We chose this architecture because it allows reward shaping while avoiding re-training the bot every time we design a new reward function. It also allows us to assign a reward to states in the environment, as well as actions of the bot. 

\subsection{Policy Dataset}
We create distinctive playtsyles for the policy dataset through reward shaping (RS). Thus, we manually design a particular reward function so that the resulting policy has a unique playtyle. Finally, we store all policies in our policy dataset, and for each one, we also compute its playstyle similarity metrics. 

\subsection{Playstyle Similarity Metric}
\label{subsec::similaritymetric}
The playstyle similarity metric is designed to characterize, in a quantitative manner, the playstyle of each policy. To do so, we keep track of how frequently each action is used and the visiting frequency of each state in the environment. 

\subsection{Search Function}
Our framework provides a function that allows users to search for policies in our dataset that exhibit the desired playstyle. This search function can use either gameplay traces from the user or a policy in our dataset as input. For the first option, users take control of the character in the game and demonstrate the desired playstyle. Once the user ends the gameplay demonstration, our framework characterizes the playstyle and provides the policy in our dataset that most closely resembles the exhibited gameplay plan as output. The second option for searching takes one of the policies in our dataset as input; users only need to select the preferred policy/plasytyle as input, and our framework will provide the closest playstyle to this input in our dataset. To compute the similarity between playstyles, we use mean-squared error to estimate the differences between playstyle metrics.

\subsection{Delimiting the Activation of Policies}
The end-user should be able to delimit the activation of each policy, which shapes the entire playstyle of the bot; depending on the game, this procedure may vary. For instance, in a platforming game, we could delimit the activation of a policy to a determined segment of a game level. However, for a fighting game, we could assign a specific policy according to the health level of the bot. 

\section{MarioMix}
\label{sec::MarioMix}
Based on the generic framework we proposed in the previous section, we designed the authoring tool called MarioMix. This tool is aimed at creating aligned behaviors for the game \textit{Super Mario Bros.}, where end-users are game designers. 

\subsection{Interface}
\label{sec::Interface}

\begin{figure*}[!ht]
\begin{center}
\includegraphics[width = 0.9 \textwidth ]{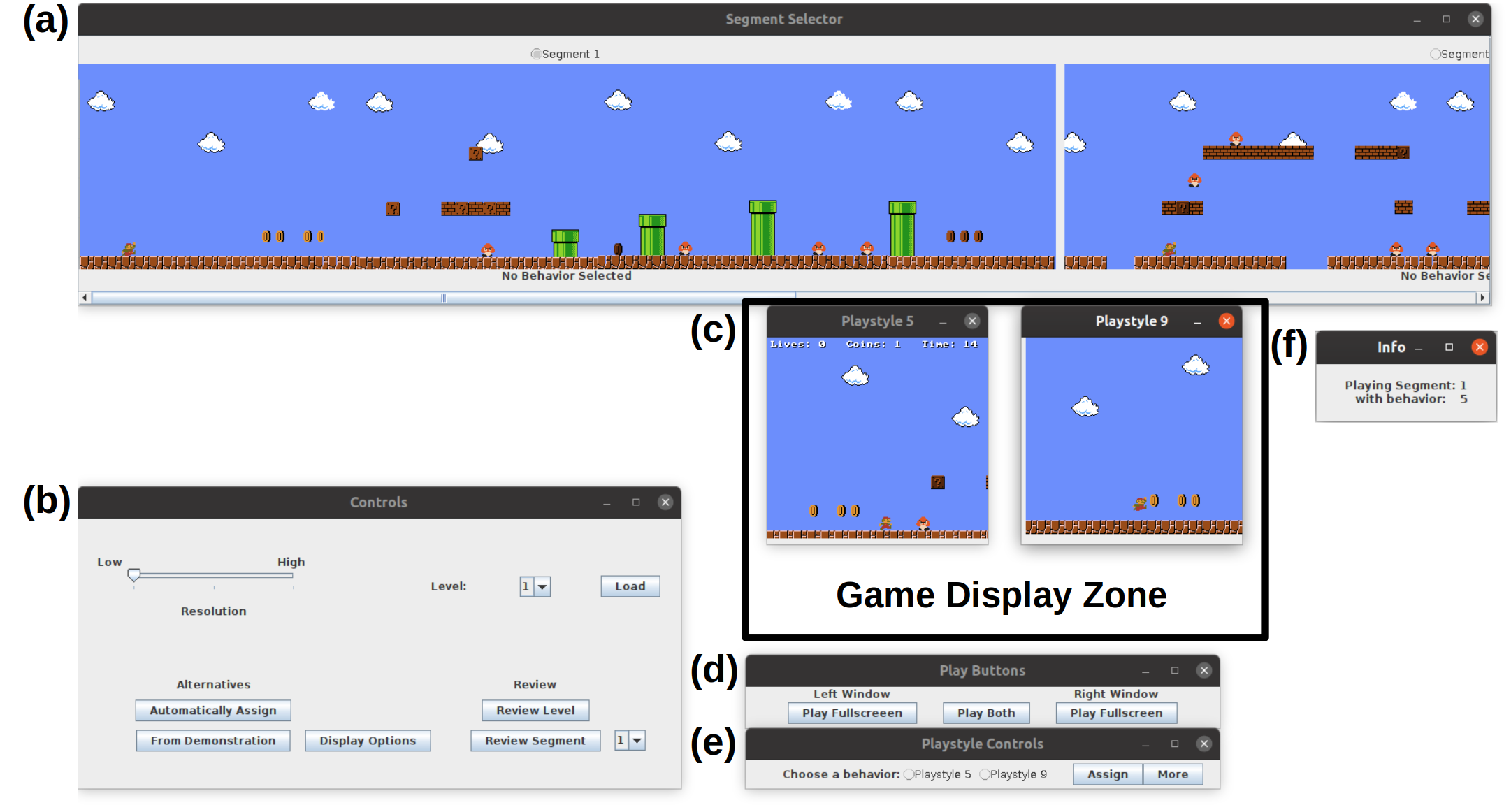}
\caption{The MarioMix interface.}
\label{Fig:mainInterface}
\end{center}
\end{figure*}

In Figure \ref{Fig:mainInterface}, we display the MarioMix interface. We will introduce its elements trough an example of a user study, which we explain in greater detail in Section \ref{sec::UserStudy}. 

In this example, a game developer named Sophia wants to create a synthetic tester for \textit{Super Mario Bros.} using MarioMix. First, she selects the desired game level and the resolution of the segments using the corresponding controls in panel \textbf{(b)}. Sophia can select the game level to display by using the drop-down list and ``load'' button on the top-right corner of the panel. Furthermore, she can change the resolution of the current game level with the slider on the top left corner of the panel. This slider has three different positions, and it controls the number of segments in a game level. We segment the game levels in $3$, $5$, and $10$ parts for low, medium, and high resolutions, respectively. 

Now Sophia can focus on assigning playstyles to all the segments in the game level that are presented in panel \textbf{(a)}. From this panel, she can select a segment to work with by selecting the corresponding checkbox on top of each segment thumbnail. Below each thumbnail, the name of the assigned playstyle is presented. This element also provides Sophia with an overview of the current progress on designing a mixture of playstyles for the whole game level at hand.

The next step is finding the right playstyle for each segment. To do so, Sophia begins by reviewing two proposed playstyles, which, in the first iteration, are presented at random by MarioMix, using the buttons in panel \textbf{(d)}. By pressing the button ``play fullscreen'' on the left side of the panel, only the playstyle option on the left is displayed using all the space in \textbf{(c)}; the function of ``play fullscreen'' on the right is similar but for the playstyle option on the right. On the example in Figure \ref{Fig:mainInterface}, the left and right behavior options are playstyles $5$ and $9$, respectively. The button in the center, ``play both,'' displays both playstyle options simultaneously.

However, Sophia did not find the type of playstyle she wanted in either playstyle $5$ or $9$. Hence, she decides to explore the playstyle space using her own gameplay as input. She starts this process by pressing the ``from demonstration'' button in panel \textbf{(b)}. After pressing this button, a window with the game is shown in the \textit{game display zone} \textbf{(f)}, with the game character in the start position of the current game level. In this game window, Sophia has control of the main character, Mario. Once the character dies, or Sophia decides to close the game window, her playstyle metrics are computed. Then, by pressing the ``display options'' button in panel \textbf{(b)}, MarioMix presents Sophia (on the \textit{game display zone}) the two most similar playstyles that match her exhibited behavior.

This time, although Sophia likes one of the playstyles proposed by MarioMix, she thinks it not quite what she wants, so she continues exploring the plastyle space using this suggested playtyle behavior as input. To do so, she first selects the input behavior using the checkboxes in panel \textbf{(e)}, and then she presses the ``more'' button in the same panel. After doing this, MarioMix searches for a different behavior that is similar to the one selected in the checkboxes and displays it in the \textit{game display zone}. This time, Sohpia finds the perfect playtyle, so she selects in the checkboxes and then presses the ``assign'' button. 

After assigning the perfect playstyles for the segments that Sophia considers the most important, she is ready to see how Mario plays the whole game level. Therefore, she uses the automatically assign option so that MarioMix finishes her job. When the ``automatically assign'' button is pressed, MarioMix automatically assigns a behavior to segments without an assigned behavior or marked as ``no behavior selected.'' The heuristic used to select a behavior starts from the first segment and repeats the selection of behavior performed by Sophia to unassigned segments. If a segment has an assigned playstyle, MarioMix randomly selects a playstyle from the dataset and assigns it to the whole game level at hand.

Once Sophia finishes assigning plasytyles to all the segments in the game level, she decides to review how Mario plays. The tools used for reviewing the assigned playstyles are two buttons and the drop-down menu on the bottom-right corner of panel \textbf{(b)}. When Sophia presses the ``review level'' button, MarioMix displays the bot playing with the designated playstyles on the \textit{game display zone}. Panel \textbf{(f)} also informs her of the current segment and playstyle that the bot is playing. However, if Sophia prefers to review only a particular segment, she can do so by selecting the ``review segment'' button and the drop-down menu next to it; if she clicks on this button, MarioMix displays the bot playing the selected segment in the drop-down menu.

\section{Bot Definition}
\label{sec::BotDefinition}
In this section, we explain how we discretized the game \textit{Super Mario Bros.} for MarioMix. Additionally, we present the handcrafted reward functions used to regenerate varied playstyles.

\subsection{State Representation}
\begin{figure}[!ht]
  \begin{center}
  \includegraphics[width = 0.9 \columnwidth ]{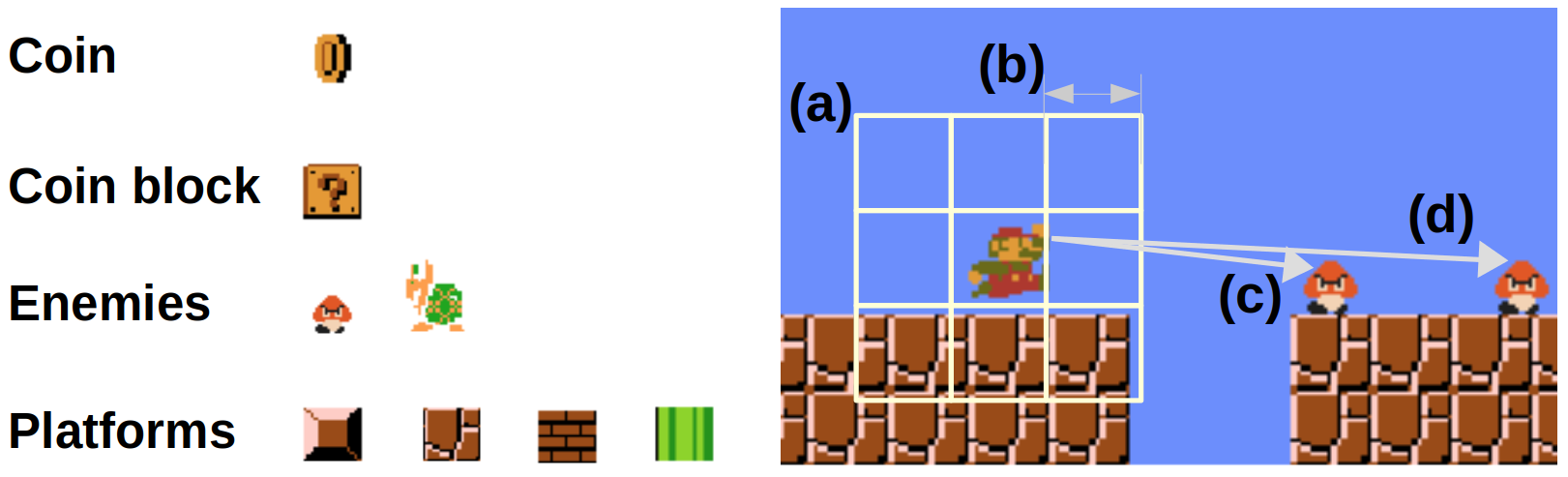}
  \caption{The state representation of our bot in MarioMix.}
  \label{Fig:genome}
  \end{center}
  \end{figure}

For the representation of \textit{Super Mario Bros.} (see Figure \ref{Fig:genome}) as a Markov decision process (MDP), we use a $3 \times 3$ grid of variables \textbf{(a)} that code terrain information. This grid can recognize platforms, coin blocks, and coins. Our MDP also accounts for the position (in $X$ and $Y$ axes) of the two enemies, \textbf{(c)} and \textbf{(d)}, closest to Mario. The position variables are discretized into $7$ values that range from $0$ to $6$. We also added a binary variable \textbf{(b)} that detects whether there is a cliff in front of Mario.

Our bot can perform $10$ different actions. Mario performs the next actions for both the left and right side of the screen: the actions walk, run, jump, and quick jump. Furthermore, Mario can do nothing, as well as perform a neutral jump.

\subsection{Policy Dataset}
\label{subsec::RewardFunctions}
To create multiple playstyles for our policy dataset, we hand-engineered $11$ reward functions (RFs). These RFs are intended to generate a wide range of playstyles within the range allowed by our discretization.

First, we focus on creating policies that maximize the next behavior traits in the bot: the speed at which a level is completed, how it overcomes enemies, how it approaches coins, and how often it jumps. For the speed trait, we designed RFs that create bots that advance either very quickly or slowly in the level. Regarding how the bot overcomes enemies, we crafted RFs that made the bot very cautious about enemies (i.e., it avoided them as much as possible) or highly likely to kill them. Additionally, we designed RFs that create bots that either collect as many coins as possible or ignore them. Finally, for the jump preference, some bots jump as much as possible, while others avoid jumping. With the aforementioned base playstyle as a foundation, we combined them to create $11$ unique bot behaviors for MarioMix.

\section{User Study}
\label{sec::UserStudy}
For the MarioMix user study, two expert game developers employed in the game industry and lacking knowledge of academic machine learning were asked to create bot behaviors for three game levels of \textit{Super Mario Bros.} Given the current situation regarding COVID-19, these expert users tested MarioMix remotely. After completing the given task, we asked the experts a few questions about the usability of MarioMix via an online survey. Additionally, we asked for more feedback via phone and e-mails.

\subsection{Task}
We presented MarioMix to the experts as a tool to design varied behaviors for synthetic testers or bots that are intended to playtest a game automatically. Therefore, the experts focused on crafting a mixture of behaviors intended for use in playtesting \textit{Super Mario Bros.} They paid close attention to how the bot plays in specific situations, such as how the bot manages to survive a specific group of enemies in the level. The expert users were tasked with creating a mixture of behaviors for three different game levels in MarioMix.

\subsection{Survey}
Once the experts completed the assigned task, they took a survey containing the following closed-ended questions:

\begin{enumerate}
  \item Have you ever played the game called \textit{Super Mario Bros.}? Options: Yes, No, I've only watched other people play it.
  
  \item Would you like to have a tool like MarioMix for the current game you are developing? Options: yes, no.

  \item How close are the bot behaviors you created to what you had envisioned? Options: not similar at all, some resemblance, very similar, a perfect match.
\end{enumerate}

Then, we asked the expert users open-ended questions to better understand their most effective workflow while using MarioMix:

\begin{enumerate}
  \item How would you describe the most effective workflow using MarioMix?
  
  \item What was your preferred resolution, and why?

  \item Which way of watching behavior options do you prefer? (two at the same time or one by one) 
\end{enumerate}

Finally, we talked with the experts to obtain more general feedback about the usability of MarioMix in an industry context.

\section{Results}
\label{sec::Results}
Overall, the participants in our user study responded positively to the MarioMix interaction method of creating bot behavior. Both experts agreed that the interaction procedure of MarioMix is effective at creating bot behaviors that align with their intentions. Moreover, the participants found the evaluation procedure entertaining, as well as an interesting and effective way to evaluate bot playstyle by observing short gameplay clips.

The participants spent between $40$ and $60$ minutes using MarioMix. First, we provided a short demonstration of the capabilities of MarioMix and explained the task they were to perform. After they completed the task, we conducted the survey.

Regarding the closed-ended questions in our survey, \textbf{(1)} all participants indicated they had played \textit{Super Mario Bros.} previously, and \textbf{(2)} they would like to have a tool like MarioMix to create bot behaviors for their games. When asked \textbf{(3)} how close the final behaviors were to what they had envisioned, their answers were either ``some resemblance'' or ``very similar.'' 

Their responses to open-ended question \textbf{(1)} revealed that both participants had a similar method of using MarioMix. Initially, participants tried each of the different resolutions, and after a couple of minutes of testing, they chose one and kept it for the remainder of the task. Then, they tried each of the behaviors suggested by default in different segments, and when they found they did not like any of the playstyles at hand for a particular segment, they explored the dataset using the ``more'' function. This exploration procedure continued for a few iterations, and when they did not find a compelling playstyle, they changed the search procedure to the ``from demonstration'' approach. Once they assigned all segments a playstyle, they reviewed the whole game level using the ``review level'' function. If they found a problem in a segment, they found a new playstyle that was a better fit for that part of the game level.

On question \textbf{(2)} in the open-ended section, both participants preferred to use a medium resolution ($5$ segments per game level). The experts mentioned that the high-resolution segments are too short for a correct assessment of the playstyles, while the low resolution does not provide enough control to make the bot play as they intended in certain areas of the game level. From low to high, the mean duration of gameplay clips were $2.5$, $5.2$, and $8.7$ seconds, respectively.

Their answers to the open-ended question \textbf{(3)} reveal that playing both behavior options side by side is the most effective for a quick preference evaluation. Participants reported that by watching both playstyles simultaneously, they could get a broad understanding of the behaviors and how they compared to each other, which made it easy to spot unwanted traits. However, when participants were looking for a particular feature in the behavior, examining playstyles individually was more effective. The participants typically performed a coarse assessment by watching both playstyles simultaneously and then refined the search by watching playstyles individually.

In their feedback regarding the interface, the experts stated they would like to have more control over the creation of segments. In particular, they mentioned that they would like to be able to manually create the segments so that they can give more resolution to the parts of the game level they consider most relevant.

It is worth mentioning that participants also found the ``more'' option compelling because it allowed them to find playstyles that they had not previously considered appropriate for performing tests. Thus, this kind of exploration provides users a wider view of possible ways to play the game at hand.

Moreover, the experts pointed out that the ``from demonstration'' approach is useful in most cases since it helped them find the required playstyle. However, it did not perform well when participants were looking for a precise gameplay trace. For instance, one participant stated, ``I'd like Mario to kill the Koopa Troopa and then hit the shell to kill the remaining enemies.'' Since our playstyle similarity metric characterizes the character's behavior in a more general manner, we cannot incorporate these kinds of specific events into gameplay.

\section{Discussion and Future Work}
\label{sec::Discussion}
From the user study, we can conclude that the top-down method of our generic framework for tackling the agent alignment problem is effective; the participants consider that the behaviors they created for synthetic testers are close enough to their intentions to be useful in a real-world scenario. Nevertheless, they would like to be able to create more precise behaviors for certain events in the game. Additionally, the small number of policies in our dataset limits the variety of plasytyles users can generate. These observations present new challenges: how to automatically create a more diverse number of policies/playstyles and how to create such accurate bot behaviors while minimizing the amount of feedback needed from users. 

The positive outcomes of the user study validate MarioMix as an effective tool that game developers who lack knowledge of academic machine learning (ML) can create effective bot behaviors. However, adapting MarioMix for different games would require an expert on ML. In future work, we would like to investigate ways to create tools for non-experts in ML and enable them to adapt tools like MarioMix for use in other games. As far as we know, this research direction is unexplored.

Although involving professional game developers in the user study increases the ecological validity of our results, we still need to perform more tests to confirm the generalization of our claims. To fully validate our generic framework, we would like to adapt more games of different genres and test them with other professionals in the game industry. However, the results of the presented preliminary evaluation are promising.

Finally, one advantage of model-based reinforcement learning is that we can easily change the reward function of a bot to adapt its behavior. We would like to explore the creation of companions \cite{Emmerich2018} that adapt according to the game developer's preferences.

\section{Conclusions}
\label{sec::Conclusions}
In this paper, we presented a generic framework for creating aligned bot behaviors. In particular, our proposed approach provides users with novel interaction procedures that allow them to create personalized bot behaviors with a top-down method, which consists of evaluating short gameplay clips.

As a proof-of-concept of our framework, we built a tool designed around it called MarioMix. This tool was evaluated by game developers working in the game industry. The preliminary results of this evaluation are positive, as the experts agree that their current workflow would benefit from a tool like MarioMix. Furthermore, these results are encouraging for the adoption of academic AI techniques in commercial games.

\section{Acknowledgements}
This work was supported by JST CREST Grant Number JPMJCR17A1, Japan.
\bibliographystyle{ACM-Reference-Format}
\bibliography{Bibliography}

\appendix



\end{document}